\documentclass{ws-procs9x6}

\begin{document}

\title{Entanglement of two individual atoms using the Rydberg blockade}

\author{A. Browaeys$^*$,  A. Ga\"{e}tan, T. Wilk, C. Evellin, J. Wolters, \\
Y. Miroshnychenko, P. Grangier}
\address{Laboratoire Charles Fabry de l'Institut d'Optique, CNRS, Univ. Paris-sud,\\
Campus Polytechnique,
RD 128, 91127 Palaiseau cedex, France\\
$^*$E-mail: antoine.browaeys@institutoptique.fr}
\author{P. Pillet, D. Comparat, A. Chotia, M. Viteau}
\address{Laboratoire Aim\'{e} Cotton, CNRS, Univ Paris-Sud, B\^{a}timent 505, \\
Campus d'Orsay,
91405 Orsay cedex, France.}

\begin{abstract}
We report on our recent progress on the manipulation of
single rubidium atoms trapped in optical tweezers and the generation
of entanglement between two atoms, each individually trapped
in neighboring tweezers. To create an entangled state
of two atoms in their ground states, we make use of the
Rydberg blockade mechanism. The degree of entanglement is measured
using global rotations of the internal states of both atoms.
Such internal state rotations on a single atom are demonstrated with a
high fidelity.
\end{abstract}

\bodymatter

\vspace{5mm}

\section{Introduction}
Entanglement
has been proposed as a ressource for quantum
information processing, for quantum metrology~\cite{Roos06}, and  for the
study of quantum correlated systems~\cite{Amico08}.
It has already been demonstrated in many systems, such as photons~\cite{Aspect82},
ions~\cite{BlattWinelandNat08}, hybrid systems composed of an atom and a photon~\cite{Blinov04},
atomic ensembles~\cite{Julsgaard01,Chou05}, and superconducting circuits~\cite{Steffen06}.
Regarding the entanglement of neutral atoms, so far two different approaches have been realized.
One method relies on the interaction of transient Rydberg atoms with a high-finesse microwave cavity
and results in entanglement of the atoms in different Rydberg states~\cite{Hagley97}.
The other approach uses s-wave collisions between ultra-cold
atoms in an optical lattice~\cite{Mandel03,Anderlini07}.
Here, we demonstrate a different approach to create entanglement of two individual
atoms where we use the strong interaction of atoms when they are in a
Rydberg state. Atoms can be excited briefly to a Rydberg state where they can interact,
and in this way their interaction can be switched on and off at will.
This approach has been proposed theoretically in the context of quantum information
processing~\cite{Jaksch00, Lukin01, Saffman05,Moller08, Mueller09} and
is in principle deterministic and scalable.

\section{Single atoms in optical tweezers}

\begin{figure}
\begin{center}
\psfig{file=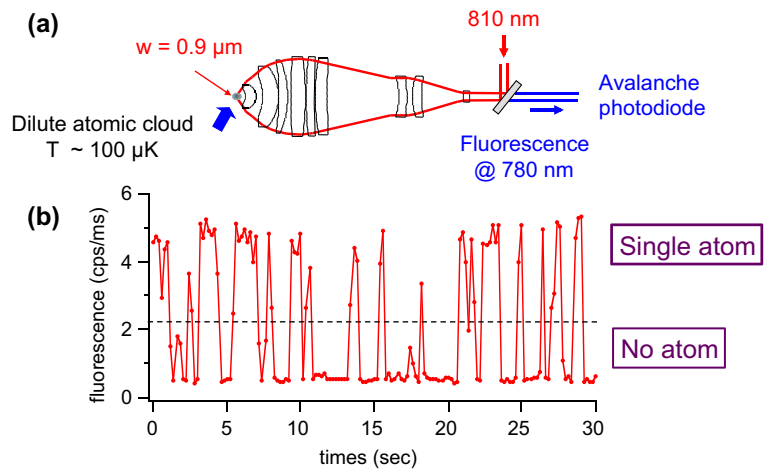,width=10cm}
\end{center}
\caption{(a) Optical setup for single-atom trapping. A homemade large numerical aperture objective
consisting of 9 lenses focuses the tweezers light at 810~nm into an optical molasses. The same lens is used to collect
the fluorescence light emitted at 780~nm by the atom. (b) Example of the fluorescence signal collected on an avalanche photodiode. The higher level of the steps in the count rate indicate that a single atom is trapped in the tweezer.  }
\label{expsetup}
\end{figure}

In our experiment we use rubidium 87 atoms.
A laser beam tightly focused down to the diffraction limit of a
large numerical aperture lens (N.A. = 0.7) forms a dipole trap which
acts as an optical tweezer, as shown in figure~\ref{expsetup}(a).
The light field at the focal point of the lens is well approximated by a
gaussian beam with a waist $w = 0.9~\mu$m. The wavelength of the trapping
laser is 810~nm, detuned by 15~nm with respect to the D1-line of rubidium at
795~nm. The trap depth is 1~mK for a power of 0.8~mW.

The dipole trap is loaded from an optical molasses. Atoms enter the
trap randomly, and are laser-cooled by the molasses beams.
We collect the fluorescence light of the atoms induced by the
cooling lasers at 780~nm with the same large numerical aperture lens onto
an avalanche photo-diode.
We observe discrete steps in the photon count rate, a lower level
associated with background light and dark counts of the detectors and
a higher level which we attribute to the presence of a single atom in the trap.
Due to the tight trapping volume, two atoms cannot
be captured at the same time in the tweezers as an inelastic
light-induced collision expels both atoms immediately~\cite{Schlosser01}.
We set a threshold to decide whether an atom is present or not.
The detection of a single atom in the trap triggers the experimental sequence
(see figure~\ref{expsetup}b).

\begin{figure}
\begin{center}
\psfig{file=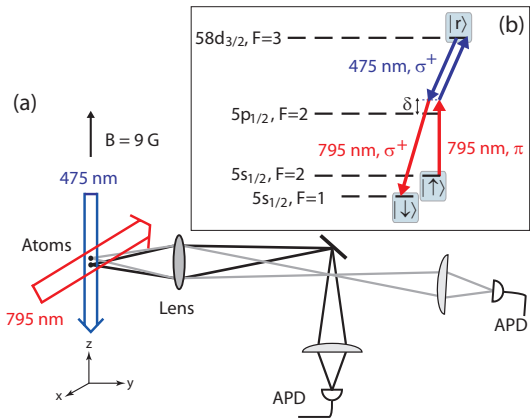,width=7cm}
\end{center}
\caption{(a) Optical setup for the collection of the light emitted by two atoms in different
dipole traps separated by 4~$\mu$m (not shown). The lasers to excite the atoms in the Rydberg states are also
shown. APD: avalanche photodiode. (b) Level structure of rubidium 87 and laser system used in the experiment.  }
\label{expsetupblocage}
\end{figure}

In order to trap two single atoms in neighboring tweezers, we send through
the same large numerical aperture lens two trapping
beams with a small angle between them. The two traps are separated by 4~$\mu$m.
Our imaging system is designed in such a way that the light coming from each
trapped atom is directed onto separate avalanche photodiodes
which allows us to discriminate for each trap whether an
atom is present or not (figure~\ref{expsetupblocage}a).
The loading of both dipole traps is random and we capture on average every
$0.5$~s an atom in each of the tweezers at the same time.

\section{Single atom internal state manipulation}\label{Raman}

We consider the two hyperfine ground states $|\!\downarrow\rangle = |F = 1, M=1\rangle$ and
$|\!\uparrow\rangle = |F = 2, M=2\rangle$ of the $5s_{1/2}$ level which are
separated by $h\times 6.8$~GHz (figure~\ref{expsetupblocage}b). We apply a
9~G magnetic field to lift the degeneracy between the Zeeman sublevels, so that $|\!\downarrow\rangle$ and
$|\!\uparrow\rangle$ form a clean two-level system.

We drive the transition between these two states using a pair of
laser beams in Raman configuration. Both lasers have a wavelength of 795~nm and
are phase-locked to each other with a frequency difference of 6.8~GHz.
They are blue detuned by 600~MHz with respect to the level $(5p_{1/2},F'=2)$.
The two beams are copropagating and focused to a waist of 130~$\mu$m.
The laser power is 40~$\mu$W in each beam resulting in a Rabi frequency of $2\pi \times
17$~MHz for each beam and a two-photon Rabi frequency $\Omega_{\uparrow\downarrow} = 2\pi\times 250$~kHz.

We measure the internal state of the atom using a push-out laser which
is tuned to the transition from $(5s_{1/2},F = 2)$ to
$(5p_{3/2},F' = 3)$. The push-out laser is applied on the atom before we check its
presence in the trap by turning back on the molasses beams and
observing the fluorescence. While an atom in state $|\!\uparrow\rangle$ will
be expelled from the trap and is absent
at the end of the sequence, an atom in state $|\!\downarrow\rangle$ will not
be influenced by the push-out laser and is still present. We note
that this method does not discriminate between Zeeman sublevels of the
$F = 1$  and $F = 2$  manifold $5s_{1/2}$.

\begin{figure}
\begin{center}
\psfig{file=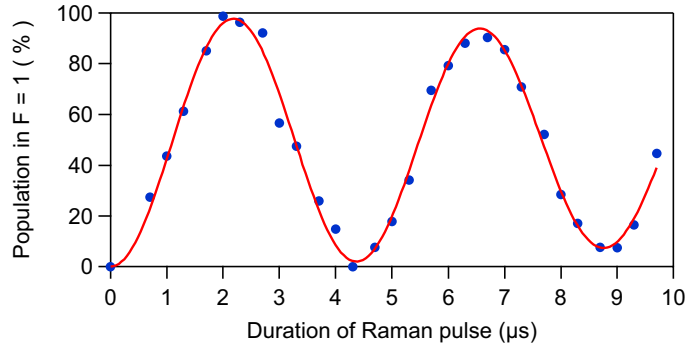,width=9cm}
\end{center}
\caption{Rabi oscillations between the states  $|\!\downarrow\rangle$
 and $|\!\uparrow\rangle$ with a Rabi frequency of 250 kHz.
 The line is a fit on the data with the model developed in reference~\cite{Jones07}. }
\label{super_Raman}
\end{figure}

To perform an internal state rotation of a single atom,
we apply the following experimental sequence. We start by pumping the atom in state
$|\!\uparrow\rangle$ by applying a 600~$\mu$s long laser pulse,
$\sigma_{+}$ polarized and tuned on the $(5s_{1/2},F = 2)$ to $(5p_{3/2},F' = 2)$
transition, together with a repumping laser tuned on the
$(5s_{1/2},F = 1)$ to $(5p_{3/2},F' = 2)$ transition.  We then apply the pair of
Raman lasers for a given duration and finally detect the atomic state with the
push-out technique. We repeat this sequence 100 times and measure the probability
to find the atom in state $|\!\downarrow\rangle$. When varying the duration of
the Raman pulse, we observe Rabi oscillations between the states $|\!\downarrow\rangle$
and $|\!\uparrow\rangle$, as shown in figure~\ref{super_Raman}.
Using the model developed in reference~\cite{Jones07}, we extract from the
contrast of the oscillation an efficiency above 99\% for the combined sequence of
preparation, rotation and detection.

\section{Rydberg blockade and entanglement}\label{blockade_entanglement}

When an atom is in a Rydberg state (principal quantum number $n\gg 1$),
one of its electrons is very far from the nucleus, typically at a distance
$n^2 a_{0}$ (with $a_{0}$ the Bohr radius). As a consequence the Rydberg atom develops
a large electric dipole moment. Two of them will therefore
interact strongly even at a distance of several micrometers.
This strong interaction can be used to prevent the simultaneous excitation
of two atoms into a Rydberg state, a mechanism known as the Rydberg blockade.

\begin{figure}
\begin{center}
\psfig{file=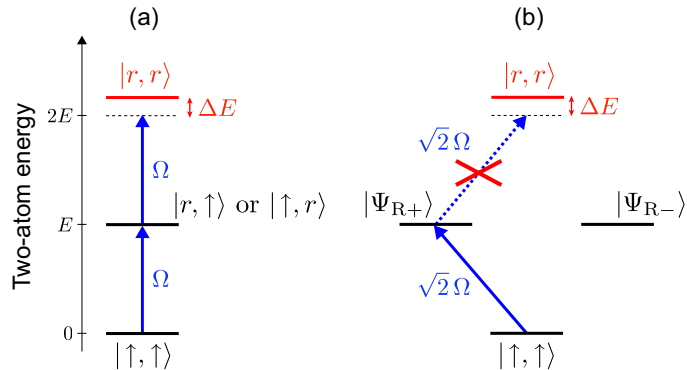,width=9cm}
\end{center}
\caption{(a) Principle of the Rydberg blockade between two atoms. (b) Principle of
the collective excitation of two atoms in the Rydberg blockade regime. }
\label{principe_blocage}
\end{figure}

The principle of the blockade is shown in figure~\ref{principe_blocage}(a).
Ground state $|\!\uparrow\rangle$ and Rydberg state $|r\rangle$
are separated by an energy $E$. The spectrum of the two-atom system
exhibits two degenerate transitions coupling $|\!\uparrow,\uparrow\rangle$
to $|\!\uparrow, r\rangle$ or $|r,\uparrow \rangle$ and these two states
to $|r,r\rangle$. However, if the atoms are close enough the energy of the doubly
excited state $|r,r\rangle$ is shifted by an amount $\Delta E$.
Then the degeneracy is lifted and a laser excitation with a linewidth smaller
than $\Delta E$ can not excite both atoms to the Rydberg state.

As a consequence of the blockade the two atoms behave collectively,
as illustrated in figure~\ref{principe_blocage}(b). If only one of the two atoms is
excited, it is convenient to use the two entangled states
$|\Psi_{{\rm R}\pm}\rangle = \frac{1}{\sqrt{2}}(e^{i {\bf k}\cdot {\bf r}_{a}}|r,\uparrow\rangle \pm
e^{i {\bf k}\cdot {\bf r}_{b}}|\!\uparrow,r\rangle)$ as a basis, where ${\bf r}_{a}$ and  ${\bf r}_{b}$
are  the positions of the two atoms, and ${\bf k}$ is related to the
wavevectors of the exciting lasers. The state $|\Psi_{{\rm R}-}\rangle$
is not coupled to the ground state, while the state $|\Psi_{{\rm R}+}\rangle$
is coupled with an effective Rabi frequency $\sqrt{2}\, \Omega$, where $\Omega$
is the Rabi frequency between $|\!\uparrow\rangle$ and $|r\rangle$ of a single atom.
In the blockade regime, where the state $|r,r\rangle$ is out of resonance,
the two atoms can be described by an effective two-level system involving
collective states $|\!\uparrow,\uparrow\rangle$ and $|\Psi_{{\rm R}+}\rangle$
coupled with a strength of $\sqrt{2}\, \Omega$. Hence, the atoms are
excited into an entangled state containing only one excited atom, with a
probability which oscillates $\sqrt{2}$ times faster than the probability
to excite one atom when it is alone.

To produce entanglement between the atoms in their ground states we start
from $|\!\uparrow,\uparrow\rangle$ and apply a pulse of duration
$\pi/(\sqrt{2}\,\Omega)$ which prepares the state $|\Psi_{{\rm R }+}\rangle$.
Then, the Rydberg state $|r\rangle$ is mapped onto the other ground state
$|\!\downarrow\rangle$ using additional lasers (wave vector $\mathbf{k'}$,
same Rabi frequency  $\Omega$) with a pulse of duration $\pi/\Omega$.
This sequence results in the maximally entangled state
\begin{equation}\label{eqno2}
|\Psi\rangle = \frac{1} {\sqrt{2}} (|\!\downarrow,\uparrow\rangle + e^{i \phi}|\!\uparrow,\downarrow\rangle),
\end{equation}
with $\phi = (\mathbf{k} -\mathbf{k'})\cdot (\mathbf{r}_{b} - \mathbf{r}_{a}) $,
assuming that the positions of the atoms are frozen during the applied pulse sequence.
If the light fields are propagating in the same direction and the energy difference between
the two ground states is small, $\mathbf{k}\simeq \mathbf{k'}$, we
deterministically generate a well defined entangled state with $\phi =0$
which is the $|\Psi^+\rangle$ Bell state.

\section{Demonstration of Rydberg blockade between two atoms and collective excitation}

We have chosen the Rydberg state $|r\rangle$=$|58d_{3/2}$,\,$F$=3,\,$M$=3$\rangle$.
The interaction energy between two atoms in this state is enhanced by
a F\"{o}rster resonance which leads to a calculated
interaction energy $\Delta E/h\approx 50$~MHz for a
distance between the atoms of 4~$\mu$m~\cite{Gaetan09}.

We excite the atoms to the Rydberg state $|r\rangle$ by a two-photon transition.
One of the excitation lasers has a wavelength of 795~nm and is detuned by several
hundreds of MHz to the blue of the transition from $|\!\uparrow\rangle$ to the
intermediate state $|5p_{1/2}, F = 2, M=2\rangle$.
The second laser has a wavelength of 474~nm and connects the
intermediate state to the Rydberg state
(see figure~\ref{expsetupblocage}b). Both laser beams illuminate the two atoms.
During the excitation ($<500$~ns), the dipole trap is turned off to avoid
an extra light-shift on the atoms. A successful excitation of an atom to the
Rydberg state is detected through the loss of the atom when the dipole trap is
turned back on, as atoms in the Rydberg state are not trapped in the tweezers.

\begin{figure}
\begin{center}
\psfig{file=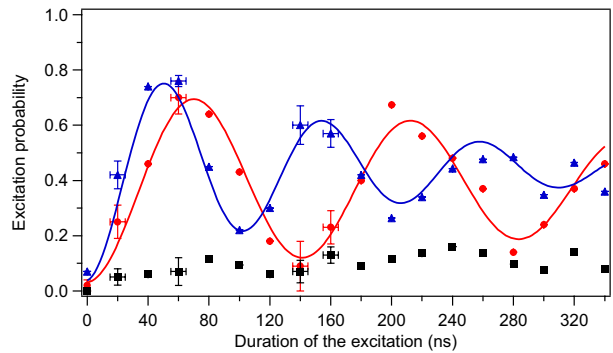,width=8cm}
\end{center}
\caption{Demonstration of the Rydberg blockade between two atoms separated by 4 $\mu$m. The dots
are the probability to excite one atom alone (second trap empty). The squares are the probability to excite
the two atoms at the same time to the Rydberg state. They show a suppression of the excitation which indicates
the blockade. The triangles are the probability to excite only one of the two atoms. It oscillates faster
than for one atom alone due to the collective behaviour. }
\label{blocage}
\end{figure}

Figure~\ref{blocage} shows the result of two experiments~\cite{Gaetan09}, where we
apply the Rydberg excitation laser pulses either to a single atom or two neighboring atoms.
In the first experiment, only one of the two dipole traps is filled with a single
atom. We prepare the atom in state $|\!\uparrow\rangle$ and send the Rydberg excitation
lasers for a given duration. Afterwards we measure if the atom is present (i.e. no
Rydberg excitation) or absent (i.e. excited to the Rydberg state). We repeat the
sequence 100 times to extract the excitation probability for a given pulse duration.
We observe Rabi oscillations between state $|\!\uparrow\rangle$ and $|r\rangle$ at
a frequency $\Omega = 2\pi\times7$~MHz. The contrast is limited in this experiment
by imperfect
optical pumping, laser intensity and frequency fluctuations
and spontaneous emission from the intermediate state.
In the second experiment, we repeat the same sequence but this time we trap an
atom in each of the two tweezers separated by 4~$\mu$m. At the end of each sequence
we measure the presence or the absence of each atom and extract the
probability to excite both atoms at the same time and the probability to
excite only one of the two atoms. Figure~\ref{blocage} shows that the
probability to excite both atoms is suppressed, as it is expected in the
blockade regime. At the same time the probability to excite only
one of the two atoms oscillates faster than the Rabi oscillation of a single atom.
The ratio of the two frequencies is 1.38, compatible with
the expected $\sqrt{2}$, and is indicative of the collective behavior
of the two atoms explained in section~\ref{blockade_entanglement}.

We note a related experimental demonstration of the blockade, complementary to
our approach~\cite{Urban09}.

\section{Entanglement of two individual atoms}

We start by preparing the two atoms in the state $|\!\uparrow,\uparrow\rangle$
and we apply the Rydberg excitation pulse of duration $\pi/(\sqrt{2}\,\Omega)$.
We then map the coherence produced between the states $|\!\uparrow\rangle$
and $|r\rangle$ onto the two hyperfine states $|\!\uparrow\rangle$ and
$|\!\downarrow\rangle$ by applying on both atoms a second pulse of duration
$\pi/\Omega$, as explained in section~\ref{blockade_entanglement}. For this
mapping, we use the same 474~nm laser and an additional laser at 795~nm, as
shown in figure~\ref{expsetupblocage}(b).

After sending the two laser pulses we measure the state of the atoms using
the push-out technique described in section~\ref{Raman}. We assign the label
$0$ when the atom is lost at the end of the sequence and the label $1$ when
it is still trapped. We measure at the end of the mapping sequence the two-atom
probabilities $P_{11}=0.06$, $P_{01}=0.34$, $P_{10}=0.31$ and $P_{00}=0.29$.
In the ideal case, the preparation of the state $|\Psi^{+}\rangle$ should lead to
$P_{11}= P_{\downarrow\downarrow}= 0$, $P_{01}=P_{\uparrow\downarrow}=1/2$,
$P_{10}= P_{\downarrow\uparrow}=1/2$ and $P_{00}=P_{\uparrow\uparrow}=0$.

The fact that $P_{00}$ is much larger than expected comes from extra losses
during the entangling sequence which we can not discriminate from atoms in state
$|\!\uparrow\rangle$ since the push-out state detection technique is also
based on atom loss. Different processes contribute to the loss
from the logical states, e.g., spontaneous emission from the intermediate
state $|5p_{1/2}, F = 2, M=2\rangle$ resulting in atoms being depumped in state
$|5s_{1/2}, F = 2, M=1\rangle$, or from atoms staying in the Rydberg state resulting
in atom loss. Intensity and frequency fluctuations of the excitation lasers
also prevent perfect excitation of the atoms. The non-zero value of $P_{11}$ is
explained by spontaneous emission from state $|5p_{1/2}, F = 2, M=2\rangle$
as well as from an imperfect blockade.

In an independent measurement we have determined the atom loss during the sequence.
Reference~\cite{Wilk09} gives more details on this study. We have measured a probability
$p \approx 0.22$ to lose one atom during the entangling sequence. This leads to
a probability to lose at least one of the two atoms
of $2p(1-p)+p^2 \approx0.39$. That means, from 100 experimental runs in average we end up
61 times with both atoms in the logical states.

\begin{figure}
\begin{center}
\psfig{file=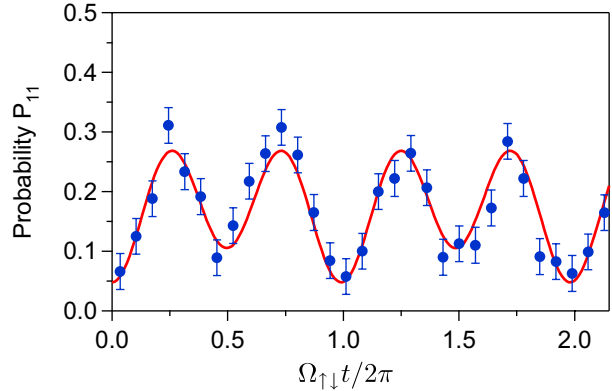,width=8cm}
\end{center}
\caption{Probability to recapture both atoms at the end of the entangling sequence
followed by a Raman rotation. We vary for each point the duration of the
Raman analyzing pulse. A detailed analysis of this data
leads to the measurement of the fidelity of the entangling operation. }
\label{figureP11}
\end{figure}

In order to analyze the amount of entanglement, we apply global Raman
rotations on the two atoms before measuring their state~\cite{Turchette98}. We vary the
duration of the Raman pulse and extract the probability $P_{11}(\Omega_{\uparrow\downarrow}t)$ as
shown in figure~\ref{figureP11}. The probability $P_{11}(\Omega_{\uparrow\downarrow}t)$ includes
only events where both atoms are recaptured at the end of
the entangling and rotation sequence, and is therefore of particular interest.
We calculate from a model $P_{11}(0) =P_{\downarrow\downarrow}$ and  $P_{11}(\pi) =P_{\uparrow\uparrow}$. A more detailed analysis of the evolution of $P_{11}(\Omega_{\uparrow\downarrow}t)$
for various Raman pulse duration is explained in reference~\cite{Wilk09}.
The average value of $P_{11}(\Omega_{\uparrow\downarrow}t)$ is related to the fidelity of the state
with respect to the expected
$|\Psi^+\rangle$ Bell state, which is defined as $F=\langle\Psi^+ |\hat\rho| \Psi^+\rangle$, with $\hat\rho$ the density matrix describing the two-atom system. From the data we extract a fidelity of the entangling sequence
$F=0.46$. This value is lower than the threshold of 0.5 which has to be overcome to prove
the quantum nature of the correlations. However, this fidelity takes into account all events,
even those for which one of the two atoms, or both, are lost from the logical states
at the end of the sequence. To retrieve an entanglement fidelity of the remaining pairs of atoms,
we calculate a renormalized fidelity of $F' = F/0.61\approx 0.75$. This value
is larger than the required threshold for a Bell's inequality test, had we a
way to post-select on the events where only pair of atoms is present.

In our experimental implementation the value of the fidelity is currently limited by
spontaneous emission, as well as laser intensity and frequency fluctuations.
The residual motion of the atoms between the two entangling pulses, which in
principle does not allow one to consider the atomic motion as frozen, as was
done in section~\ref{blockade_entanglement}, causes only a small reduction of
the observed fidelity.

\section{Conclusion}
In this paper we demonstrated our ability to manipulate the internal state of a single atom
trapped in an optical tweezer and to control the interaction between two atoms in neighboring traps.
The internal state of a single atom can be prepared with a high fidelity using Raman rotations.
The interaction between the atoms is controlled using laser excitation towards a Rydberg state
and manifest itself in the observation of the Rydberg blockade effect.
We make use of this effect to create the entanglement of two atoms in two hyperfine ground states.
Ongoing work is devoted to the improvement of the fidelity of the entangling operation.

\vspace{5mm}

{\bf Acknowledgments:}

We thank M. M\"uller, M. Barbieri, R. Blatt, D. Kielpinski and
P. Maunz for discussions and T. Puppe for assistance with the laser system.
We acknowledge support from the EU through the IP SCALA, IARPA and IFRAF.
A. G. and C. E. are supported by a DGA fellowship and Y. M. and T. W. by IFRAF.


\vspace{5mm}

\begin{thebibliography}{30}

\bibitem{Roos06} C.\,F. Roos \textit{et al.}, Nature \textbf{443}, 316 (2006).

\bibitem{Amico08} L. Amico \textit{et al.}, Rev. Mod. Phys. \textbf{80}, 517 (2008).

\bibitem{Aspect82} A. Aspect, Nature \textbf{398}, 189 (1999).

\bibitem{BlattWinelandNat08} R. Blatt, and D. Wineland, Nature Ê\textbf{453}, 1008 (2008).

\bibitem{Blinov04} B.\,B. Blinov \textit{et al.}, Nature \textbf{428}, 153 (2004).

\bibitem{Julsgaard01} B. Julsgaard, A. Kozhekin, and E.\,S. Polzik, Nature \textbf{413}, 400 (2001).

\bibitem{Chou05} C.\,W. Chou \textit{et al.}, Nature \textbf{438}, 828 (2005).

\bibitem{Steffen06} M. Steffen {\it et al.}, Science \textbf{313}, 1423 (2006).

\bibitem{Hagley97} E. Hagley {\it et al.}, Phys. Rev. Lett. \textbf{79}, 1 (1997).

\bibitem{Mandel03} O. Mandel {\it et al.}, Nature \textbf{425}, 937 (2003).

\bibitem{Anderlini07} M. Anderlini {\it et al.}, Nature \textbf{448}, 452 (2007).

\bibitem{Jaksch00} D. Jaksch {\it et al.}, Phys. Rev. Lett. \textbf{85}, 2208 (2000).

\bibitem{Lukin01} M.\,D. Lukin {\it et al.}, Phys. Rev. Lett. \textbf{87}, 037901 (2001).

\bibitem{Saffman05} M. Saffman, and T.\,G. Walker, Phys. Rev. A \textbf{72}, 022347 (2005).

\bibitem{Moller08} D. M{\o}ller {\it et al.}, Phys. Rev. Lett. \textbf{100}, 170504 (2008).

\bibitem{Mueller09} M. M\"{u}ller {\it et al.}, Phys. Rev. Lett. \textbf{102}, 170502 (2009).


\bibitem{Schlosser01} N. Schlosser {\it et al.}, Nature \textbf{411}, 1024 (2001).

\bibitem{Jones07} M.\,P.\,A. Jones {\it et al.}, Phys. Rev. A \textbf{75}, 040301 (2007).

\bibitem{Gaetan09} A. Ga\"{e}tan {\it et al.}, Nature Phys. \textbf{5}, 115 (2009).


\bibitem{Urban09} E. Urban {\it et al.}, Nature Phys. \textbf{5}, 110 (2009).


\bibitem{Wilk09} T. Wilk,  {\it et al.}, arXiv:0908.0454.

\bibitem{Turchette98} Q.\,A. Turchette {\it et al.}, Phys. Rev. Lett. \textbf{81}, 3631 (1998).

%
%
%
%


\end{thebibliography}
\end{document}